\documentclass[%
 reprint,
 amsmath,amssymb,
 aps,
]{revtex4-2}

\usepackage{graphicx}
\usepackage{dcolumn}
\usepackage{bm}

\begin{document}

\preprint{APS/123-QED}

\title{Virus Assembly Pathways inside a Host Cell}

\author{Sanaz Panahandeh}
\email{spana002@ucr.edu}
\affiliation{Department of Physics and Astronomy, University of California, Riverside, CA 92521, USA}
\author{Siyu Li}
\affiliation{Department of Physics and Astronomy, University of California, Riverside, CA 92521, USA}
\author{Bogdan Dragnea}
\affiliation{Department of Chemistry, Indiana University, Bloomington, IN 47405, USA}
\author{Roya Zandi}
\affiliation{Department of Physics and Astronomy, University of California, Riverside, CA 92521, USA}

\begin{abstract}
Simple RNA viruses self-assemble spontaneously and encapsulate their genome into a shell called the capsid. This process is mainly driven by the attractive electrostatics interaction between the positive charges on capsid proteins and the negative charges on the genome. Despite its importance and many decades of intense research, how the virus selects and packages its native RNA inside the crowded environment of a host cell cytoplasm in the presence of an abundance of non-viral RNA and other anionic polymers, has remained a mystery.  In this paper, we perform a series of simulations to monitor the growth of viral shells and find the mechanism by which cargo-coat protein interactions can impact the structure and stability of the viral shells. We show that coat protein subunits can assemble around a globular nucleic acid core by forming non-icosahedral cages, which have been recently observed in assembly experiments involving small pieces of RNA. We find that the resulting cages are strained and can easily be split into fragments along stress lines. This suggests that such metastable non-icosahedral intermediates could be easily re-assembled into the stable native icosahedral shells if the larger wild-type genome becomes available, despite the presence of myriad of non-viral RNAs.
\end{abstract}

\maketitle


\section{\label{sec:level1}Introduction}

Positive-strand RNA viruses represent the largest genetic class of viruses, including many human, animal, and plant pathogens \cite{King2011}. Their virions contain mRNA-sense single stranded (ss) RNA which is protected by a protein cage called the capsid \cite{Flint}. The capsids of more than half of the RNA virus genera have icosahedral symmetry \cite{Rao:2006ge, weiss2005armor, johnson1997quasi,twarock2019structural}. Since the spontaneous assembly of molecular constituents into a functional virion is one of the key steps in the virus life cycle, its physical and chemical bases have been intensely studied on a number of model systems. A wealth of information has emerged from \emph{in vitro} assembly studies and detailed models that explain several salient features of the virus self-assembly have been proposed \cite{Zandi2020, Hagan2021,Chen:2007b,elrad2008mechanisms}. 

For all ssRNA icosahedral viruses, genome encapsidation happens at assembly, in cytosolic compartments where the opportunity of encapsidating other, non-cognate cargo also exists \cite{Rao:2006ge}. {\it In vitro} self-assembly studies suggest that the negatively-charged RNA recruits coat proteins (CPs) through multiple electrostatic interactions with their basic residues \cite{Hu2008, Hagan2013, dong2020effect, li2017effect, Gonca2016,zhang2014topological}. The dominant role of electrostatics is supported by \textit{in vitro} encapsidation experiments with a variety of polyanionic non-cognate cargo, ranging from other nucleic acids to nanoparticles \cite{Sun2007, garmann2014assembly}. Despite a relatively good understanding of this process \cite{elrad2010encapsulation,Hagan2013,Zandi2020, Bruinsma2016, fejer2010emergent}, the origin of selectivity of cognate RNA \emph{in vivo} encapsidation has remained a mystery because of the presence of other non-genomic polyanionic species. One explanation involves certain packaging sequences which confer a higher coat protein binding affinity to cognate viral RNA \cite{stockley2013packaging, twarock2019rna}. However, it has been pointed out that such packaging sequences might not be the sole source of specificity, with other factors being likely at play \cite{Comas-Garcia2019}. Recently, in discussing experiments that entailed a combination of charge detection mass spectrometry and cryo-electron microscopy, Bond {\it et al.} suggested capsid elastic stress at assembly as a pathway selection mechanism. In their experiments, brome mosaic virus (BMV) coat proteins were found to readily form capsids  around multiple oligonucleotides. Those capsids were smaller, non-icosahedral, and  less stable than the native ones. Based on these findings, Bond {\it et al.} speculated that shells do form around small RNAs and other polyanionic cargo but might easily fall apart, with their fragments re-assembling around cognate RNAs when they become available. This hypothesis suggests a narrowing of the possible virus assembly pathways, which would actually take advantage of the presence of non-cognate polyanions \cite{bond2020virus}. To test theoretically the possibility of this hypothesis, we have modeled capsid growth under the assumption of an elastic lattice formed of subunits having a preferred radius of curvature. 

\begin{figure} 
   \centering
    \includegraphics[width=\linewidth]{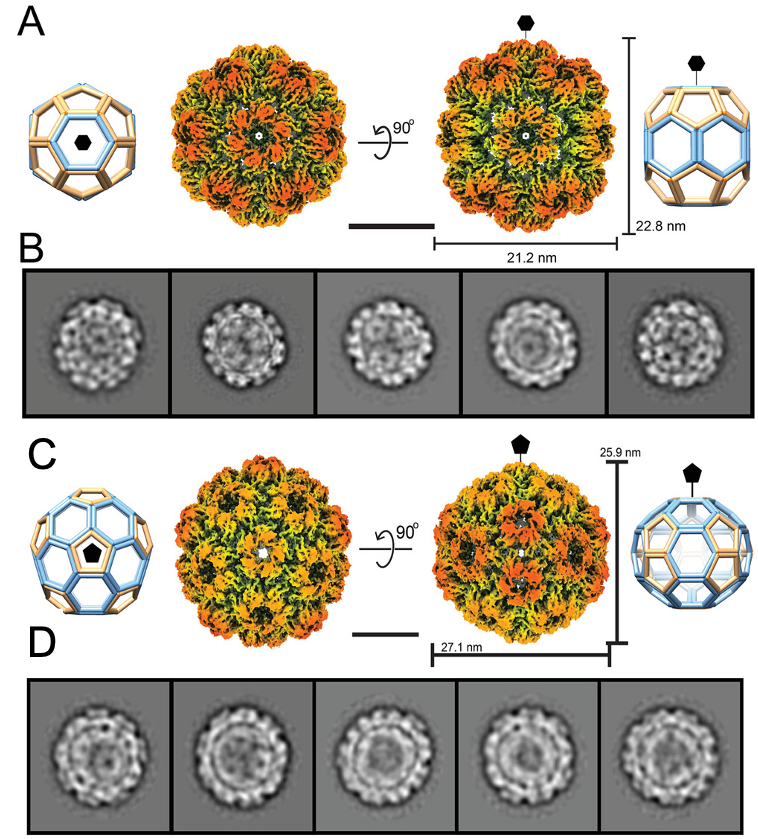} 
    \caption{\footnotesize Cryo‐EM of $E$ and $F^*$ structures with $D_{6h}$ and $D_{5h}$ symmetries obtained in the experiments of Ref.~\cite{bond2020virus}. Note that in \cite{bond2020virus} the structures $E$ and $F^*$ are called $H8$ and $H15$, respectively. A) Isosurface views of the $E$ structure at different angles. The axis of sixfold symmetry is marked. Scale bar is 10 \begin{math}\mathrm{nm}\end{math}. B) Selected reference‐free 2D class averages indicate different orientations of the particle.  C) Isosurface views of $F^*$ structure at different angles. The fivefold symmetry axis is marked. D) Selected reference‐free 2D class averages show different orientations of the particle. Scale bar is 10 \begin{math}\mathrm{nm}\end{math}. Adapted with permission from \cite{bond2020virus}.}
    \label{experiments}
\end{figure}

According to the quasi-equivalence principle, the number of proteins forming icosahedral shells is $60T$ where the $T$ number is a structural index for viral shells and is equal to $T = h^2 + k^2 + kh$ with h and k nonnegative integers \cite{CASPAR1962}. Previous {\it in vitro} assembly studies of small icosahedral ($T=3$) viruses found that capsid proteins can assemble {\it in vitro} around a variety of anionic cargoes to form isometric shells \cite{Garmann2015,Comas-Garcia2019}. In the absence of high-resolution structural or stoichiometric data, it was assumed that even if the size of the shells changes because of the size of their cargoes, the symmetry of the capsids is still the same as that of the native shell, {\it i.e.}, there are usually 12 pentamers, sitting at the vertices of an icosahedron, in a shell built otherwise from hexamers. The experimental results of Bond {\it et al.} noted above were unexpected as they found that nucleic acid oligomers and capsids proteins form structures with $D_{6h}$ and $D_{5h}$ symmetries \cite{bond2020virus}. Figure \ref{experiments} shows the cryo-EM images of two different structures obtained in Ref.~\cite{bond2020virus}.

In this paper, we study the range of elastic and geometric subunit parameters for which strained, lower-symmetry non-icosahedral structures similar to those reported by Bond {\it et al.} form.  We calculate the elastic stress distribution across these shells and discover that they should be less stable than the wild type $T=3$ shells, being strained. We also show that one of the low symmetry structures (Fig.~\ref{experiments}C) constitutes the minimum energy structure when smaller size cargoes are encapsulated, while the other one (Fig.~\ref{experiments}A) could be obtained only through non-equilibrium simulations \cite{bond2020virus}. Our theoretical study thus confirms the possibility of a pathway for the packaging process inside cells whereby capsid proteins may start assemble around non-cognate small RNAs in the cell during infection, forming shells that are strained and less stable. 

Since there is no systematic study about the impact of RNA on the capsid size and morphology as observed in the above experiments, in this paper we also explore under what conditions
 RNA can change the size and structure of viral shells.  We find that the interplay of the mechanical properties of capsid proteins and their interaction with genome is quite intriguing and that RNA can significantly modify the symmetry of viral shells giving rise to the formation of structures with different symmetries that were not found in the previous theoretical studies on empty capsids. 

This work lays out the systematic comparison of theory and experiments, which will allow us to gain a better understanding of the role of RNA in the capsid assembly pathway, stability and structure, both {\it in vivo} and {\it in vitro}.  A deeper understanding of the role of genome in virus assembly mechanisms could lead to the design principles for alternative antivirals and facilitate the fabrication and design of precise synthetic nano-structures.

\section{\label{result}Results and Discussion}

\begin{figure} 
   \centering
    \includegraphics[width=\linewidth]{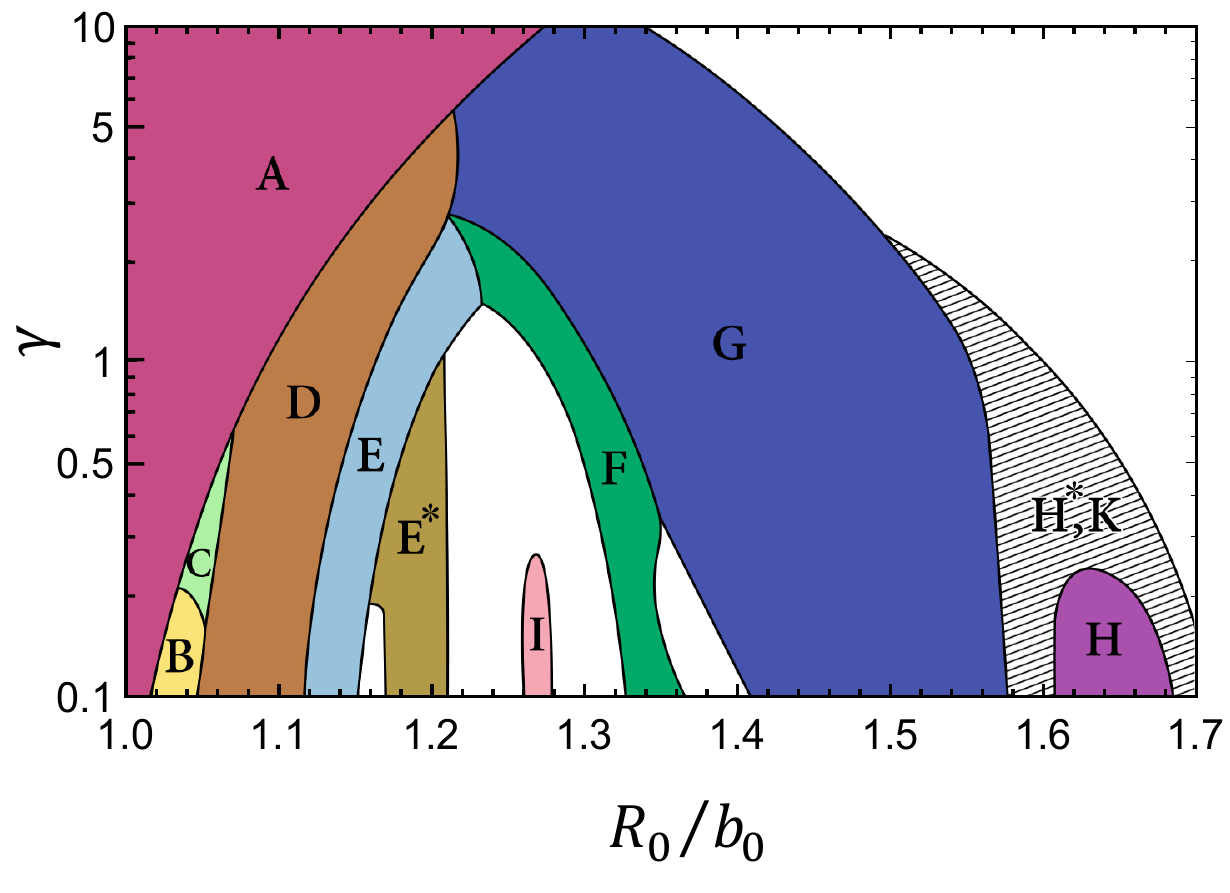} 
    \caption{\footnotesize Phase diagram of the shells assembled for different values of $\gamma$ and $R_0/b_0$ without the presence of core using the deterministic model. Each shaded area corresponds to a single type of structure that forms in that region. Shell structures with their corresponding symmetries are shown in Fig.~\ref{shape}. In the hashed regions, in addition to symmetric shells, a number of similarly sized non-symmetric structures form. The white areas correspond to regions in which different types of shells without any specific symmetry are grown. Adapted with permission from \cite{wagner2015robust}.}
    \label{empty}
\end{figure}

\begin{figure} 
\centering
\includegraphics[width=\linewidth]{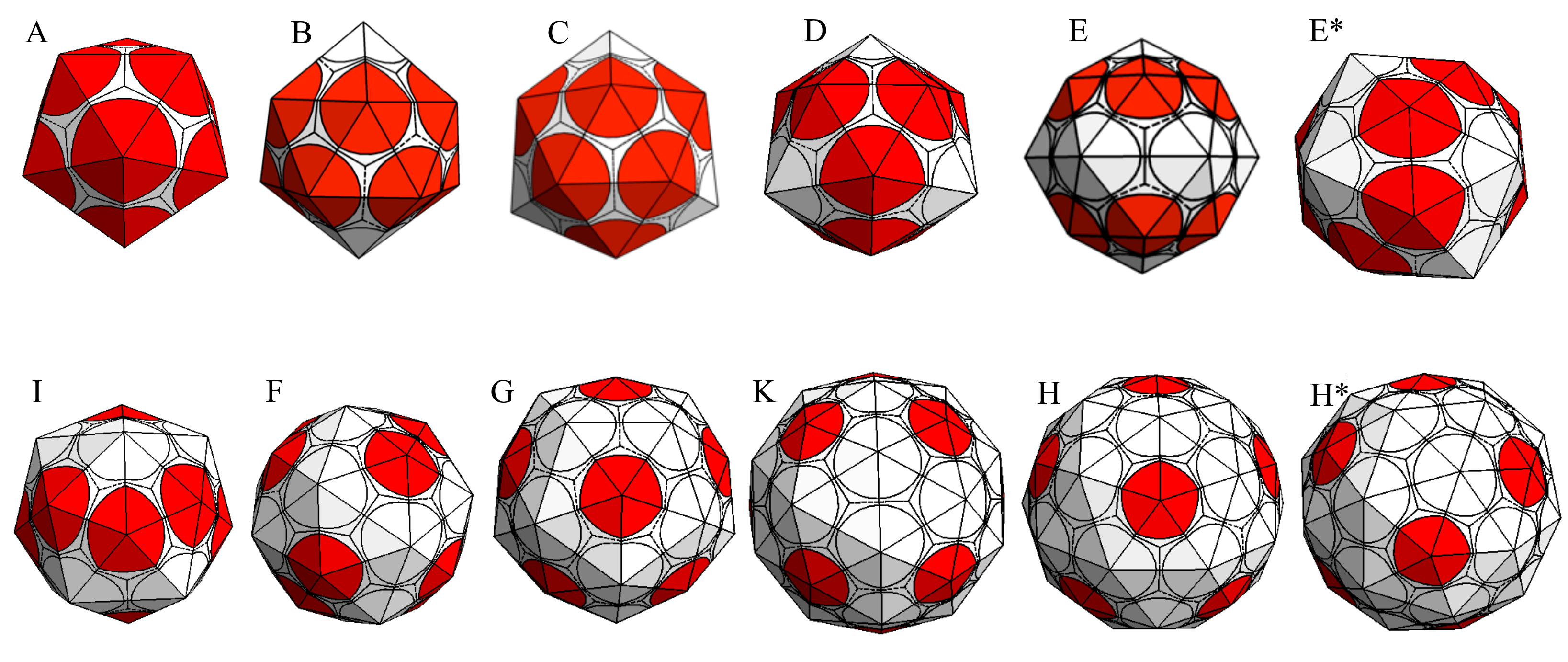} 
\caption{\footnotesize The structures obtained in the deterministic simulations corresponding to the labeled regions of the phase diagram illustrated in Fig.~\ref{empty}. The shells from left to right and top to bottom have 20, 24, 26, 28, 36, 36, 40, 50, 60, 76, 80 and 80 subunits or 12, 14, 15, 16, 20, 20, 22, 27, 32, 40, 42 and 42 capsomers respectively. Symmetries are icosahedral ($T = 1$), $D_{6}$, $D_{3}$, tetrahedral, $D_{6h}$, $D_{2}$ (tennis ball), $D_{2}$, $D_{3}$, icosahedral ($T = 3$), tetrahedral, icosahedral ($T = 4$) and $D_{5h}$, respectively. Red and white colors correspond to the location of pentamers and hexamers respectively. Adapted with permission from \cite{wagner2015robust}.}
 \label{shape}
\end{figure}

To explore the impact of RNA on the growth pathway, morphology and stability of viral shells, we construct a ``structure'' phase diagram as a function of the protein spontaneous curvature $R_0/b_0$ and the FvK number $\gamma$, for various genome size and the strength of RNA-protein interactions. We seek to determine how the phase diagram of genome-containing particles differs from the phase diagram of the empty shells obtained in Refs.~\cite{wagner2015robust,panahandeh2018equilibrium} in the absence of RNA. Figure \ref{empty} shows the results of the deterministic simulations as a two dimensional phase diagram as a function of $R_0/b_0$ and $\gamma$ in the absence of genome. Each color in the figures corresponds to a specific symmetric structure, see Fig.~\ref{shape}. 

Figure \ref{empty} clearly indicates that at higher $\gamma$s, {\it i.e.}, for the shells that are easier to bend than to stretch, we observe only a transition from $T=1$ (structure $A$) to $T=3$ (structure $G$) as a function of $R_0/b_0$. However, at lower $\gamma$s, in addition to $T=1$ and $T=3$ shells, several structures with other symmetries form as $R_0/b_0$ varies. All structures have 12 pentagons, even the aberrant ones assembled in the white region of the figure. It is notable that some of the symmetric structures ($D$, $E$, $E^*$ and $G$) in Fig.~\ref{empty} have been observed in other biological systems such as clathrin vesicles \cite{fotin2004molecular,cheng2007cryo}. 

As discussed in the method section, to make our calculations more tractable while keeping the important physical features of the system, we model the genome as a spherical core that interacts attractively with the protein subunits through LJ potential. During the simulations, we have considered that the attractive interaction between protein subunits and the genome is uniform and does not depend on the specific sequence of RNA. However, at least for certain viruses, the binding affinity is thought to vary along the genome with segments containing the packaging signals showing higher affinity than the rest \cite{twarock2018modelling}. This is important for the initial stages, but it is not clear that that the packaging signals are still playing a role after the initial nucleo-protein complex has formed and the genome has assumed a globular shape. Since the focus of the simulations is on the assembly process after the genome has assumed the globular shape, and not on the kinetics, the simplifying assumption of a spherical core is reasonable as confirmed by the fact that we obtain the same structures as observed in the experiments. When examining the influence of the core on the particle phase diagram, it is particularly interesting to focus on particle sizes that are compatible with $T=1$ and $T=3$ structures. Note that $T=1$ was observed in the {\it in vitro} experiments with Bromo Mosaic Virus (BMV), whose native morphology is a $T=3$ structure \cite{Lucas2001}. 

\begin{figure}
 \centering
  \includegraphics[width=0.7\linewidth]{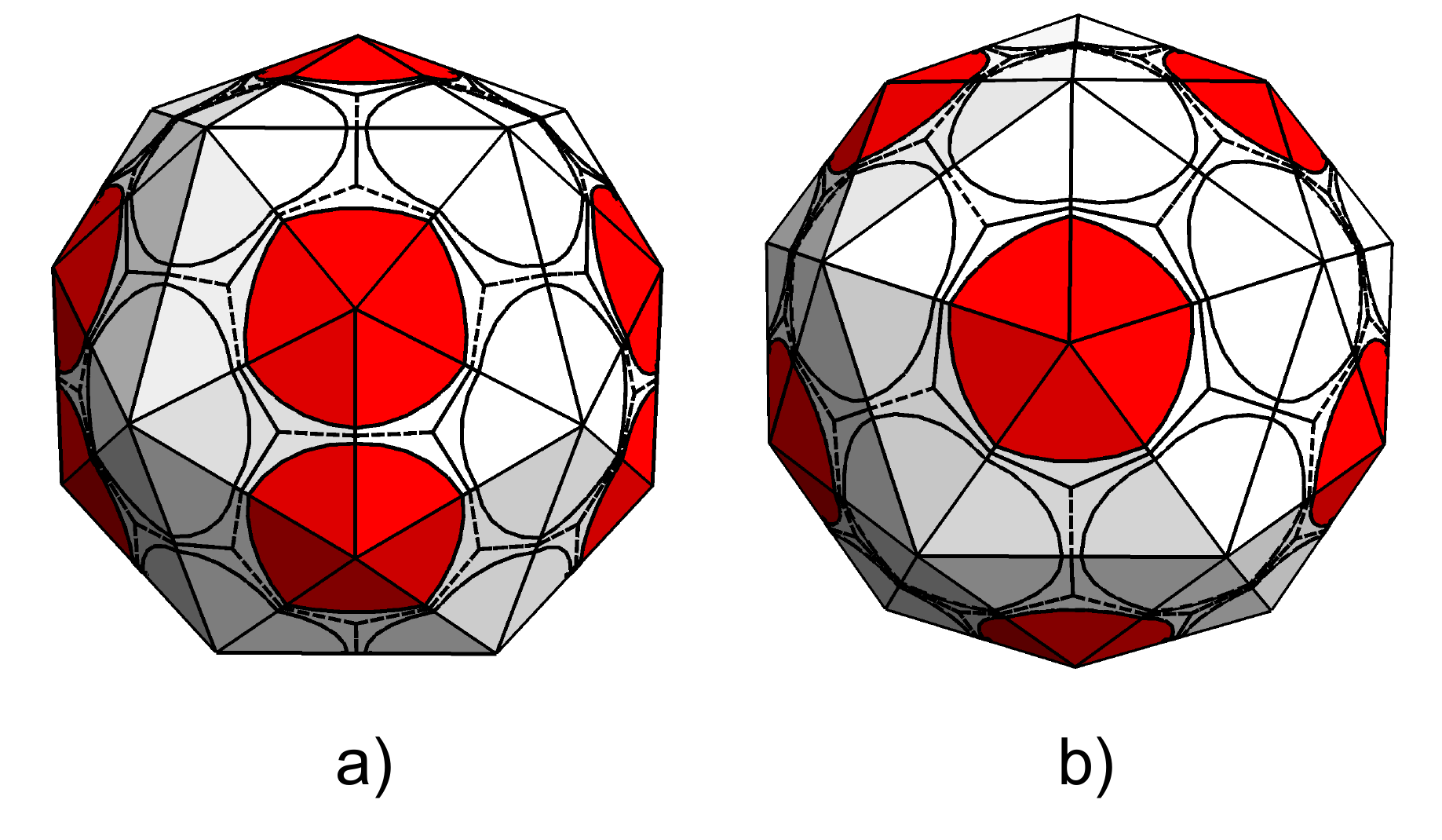}
  \caption{\footnotesize Structure $F^*$ with $D_{5h}$ symmetry has 50 subunits and 27 capsomers. a) Side view of $F^*$. It has five pairs of pentamers on the side, one single pentamer at the top and another one at the bottom. b) Top view of the $F^*$ structure.}
  \label{Fframes}
\end{figure}

We consider four different core radii $R_c/b_0= 0.95, 1.2, 1.4, 1.5$ that commensurate with the size of $T=1$, ``$E$ and $E^*$'', ``$F$ and $F^*$'', and $T=3$ structures, respectively. Note that $E$ and $E^*$ have the same size, as it is the case for $F$ and $F^*$.  We also monitor the assembly of capsids around a core with the radius $R_c/b_0=1.3$ that is not compatible with the size of any of the symmetric shells obtained in Figs.~\ref{empty} and \ref{shape}. We emphasize that $F^*$ (see Fig.~\ref{Fframes}) is exactly the structure that has been discovered in the recent self-assembly experiments of Bond {\it et al.}~\cite{bond2020virus} (see Fig.~\ref{experiments}C) and is of particular interest to our investigation of the role of genome in the structure of capsids and the virus assembly pathway in the presence of non-native RNAs. One of the interesting results of this paper is that the presence of the core is essential for the formation of $F^*$, which did not appear in the previous simulations of equilibrium \cite{panahandeh2018equilibrium} and non-equilibrium \cite{wagner2015robust} structures for empty shells.

\begin{figure}
 \centering
  \includegraphics[width=\linewidth]{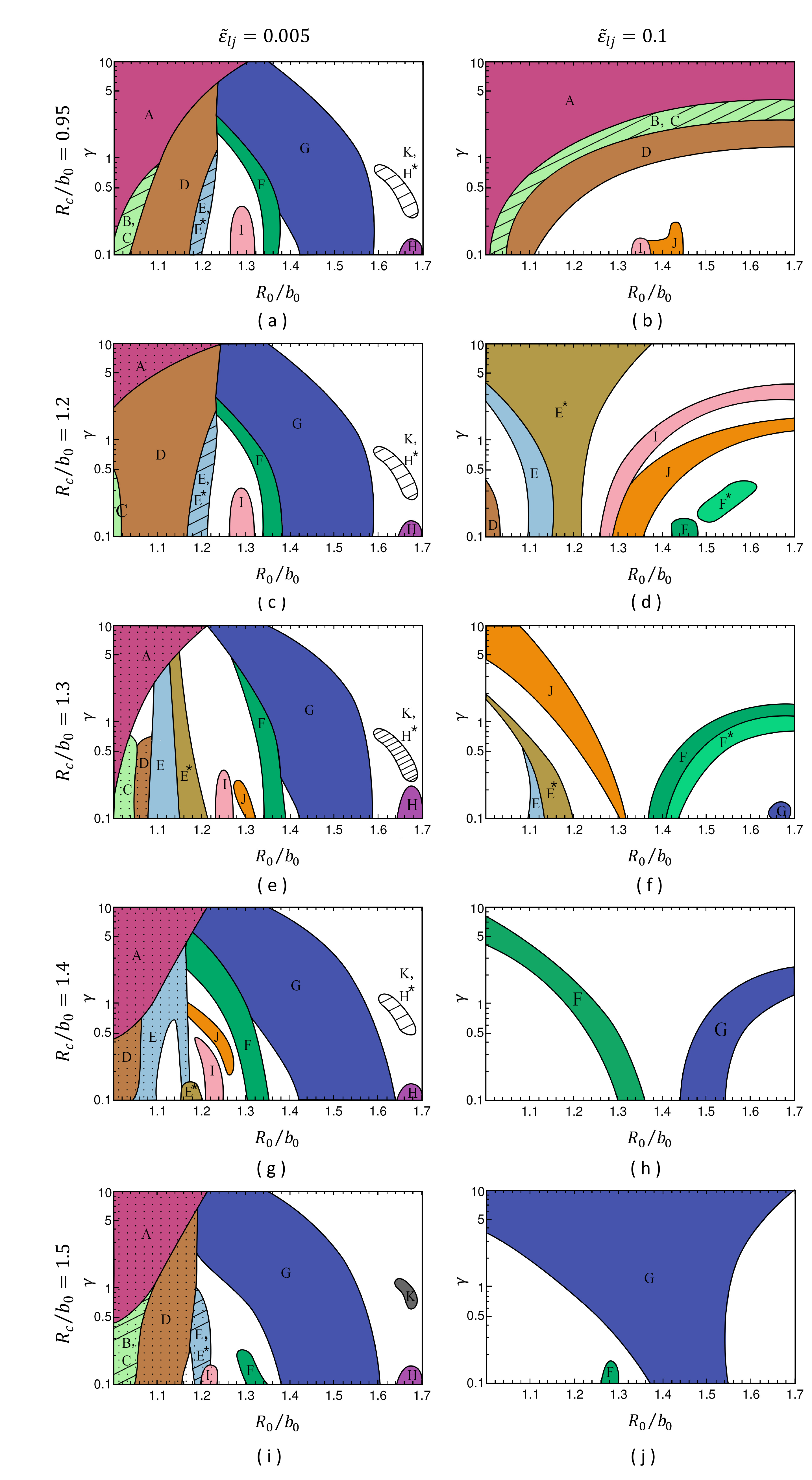}
  \caption{\footnotesize Phase diagrams of the structures obtained in the deterministic simulations in the presence of genome with five various sizes and two different strengths of the protein-genome attractive interactions. From the top to the bottom row, the core sizes are $R_c/b_0 = 0.95, 1.2, 1.3, 1.4, 1.5$. We note that the core sizes $R_c/b_0 = 0.95, 1.2, 1.4, 1.5$ are the optimal sizes to grow $T=1$, $E$ and $E^*$, $F$ and $T=3$ structures, respectively. The core size $R_c/b_0 = 1.3$ does not correspond to any of the symmetric structures. The left column corresponds to the weak core-protein interaction $\tilde{\epsilon}_{lj}= 0.005$ and the right column to the strong one $\tilde{\epsilon}_{lj}= 0.1$ . Each color corresponds to a structure with specific symmetry (see Fig.~\ref{shape}). Dotted regions correspond to the phases in which the capsids form without packaging the core. A careful review of the phase diagrams reveals that the icosahedral structures prevail even if both the spontaneous curvature and core size are optimal for the formation of another type of symmetric structure.}
  \label{noneqphase}
\end{figure}

Figure~\ref{noneqphase} shows the results of the simulations as two dimensional phase diagrams as a function of $\gamma$ and $R_0/b_0$. From the top to the bottom, the rows correspond to the core sizes of $R_c/b_0= 0.95, 1.2, 1.3, 1.4, 1.5$, respectively. The figure also illustrates the impact of the strength of genome-protein attractive interaction.  The left column in Fig.~\ref{noneqphase} corresponds to the structures for which the strength of LJ is $\tilde{\epsilon}_{lj}=\epsilon_{lj}/k_b= 0.005$, and the right column shows the results for stronger LJ interaction, $\tilde{\epsilon}_{lj}= 0.1$. Note that we divide $\epsilon_{lj}$ by $k_b$ to make it dimensionless.  Similar to Fig.~\ref{empty}, each color corresponds to a specific symmetric structure. In the dotted regions of the figure, the shells assemble without encapsulating the genome because the cost of its packaging is too high. 
 
 The phase diagrams on the left hand side of Fig.~\ref{noneqphase} are very similar to the phase diagram in Fig.~\ref{empty}, revealing that the elastic energy of the growing shell has the dominant contribution to the formation of the symmetric shells in the weak LJ regime ($\tilde{\epsilon}=0.005$). The simulations are designed such that if the size of the core is smaller than the spontaneous curvature of proteins, the strength of LJ potential is such that the shell will always assemble around it.  As the size of the core increases, the subunits whose spontaneous radius of curvature is smaller than the core radius will either assemble into an empty shell based on the phase diagram of Fig.~\ref{empty} for the relevant $\gamma$ and $R_0/b_0$ or into a filled shell but with a different symmetry than the empty one for the same $\gamma$ and $R_0/b_0$.  Thus, the main difference between the phase diagrams on the left hand side of Fig.~\ref{noneqphase} ($\tilde{\epsilon}_{lj}= 0.005$) and the ones for the empty shells (Fig.~\ref{empty}) arises from the size of the core.  At $\tilde{\epsilon}_{lj}= 0.005$, the elastic energy still controls the growth pathway, and as a result the transition lines between different structures move together without influencing the shape varieties.  Similar to Fig.~\ref{empty}, we observe two main structures form ($T=1$ and $T=3$) at high $\gamma$ values (Fig.~\ref{noneqphase}), while different structures form as $R_0/b_0$ changes for lower $\gamma$s.

 By contrast, if the strength of subunit-genome interaction is strong ($\tilde{\epsilon}_{lj}= 0.1$), the presence of RNA can completely change the structure of the capsids as illustrated in the right column of Fig.~\ref{noneqphase}. The tendency of trimers to aggregate on the core surface because of the LJ potential will compete with the bending energy of the shell, resulting in a considerable change in the phase diagram.  Most structures in the phase diagram then will have the shapes instructed by the core size.  For example, when the core size is $R_c/b_0 = 0.95$, the optimal size for a $T=1$ shell, Fig.~\ref{noneqphase}a shows that for the weak LJ potential, $T=1$ covers a small region in the phase diagram.  However, for the strong LJ potential the $T=1$ structure covers a very wide region of the phase diagram (Fig.~\ref{noneqphase}b). 
 
 Further, Fig.~\ref{noneqphase}b reveals that for $R_0/b_0>1.12$ at low $\gamma$s, only irregular shells without any specific symmetry form, in contrast to Fig.~\ref{noneqphase}a where various symmetric structures form.  The irregular shells have larger number of subunits compared to a $T=1$ structure in Fig.~\ref{noneqphase}b. At low $\gamma$s when the subunits are less rigid, they squeeze each other to make room for more subunits because of the strong attractive interaction between the proteins and the core.  As we keep increasing $\gamma$ at $R_c/b_0 = 0.95$,  the structures $D$, $C$, $B$ and $T=1$ start appearing, respectively.
 It is important to note that while these shells have different symmetries and number of subunits, they all have more or less the same size due to the strong attraction of the subunits to the core.  Figure \ref{noneqphase}b indicates that the largest symmetric capsid that forms for $R_c/b_0 = 0.95$ is the structure $D$ with tetrahedral symmetry and 28 triangular subunits. As $\gamma$ increases, the number of subunits decreases and $T=1$ becomes the dominant structure, regardless of the value of $R_0/b_0$. Without the core, the $T=1$ structure would not form if $R_0/b_0>1.28$. 
 
Increasing the genome size to $R_c/b_0=1.2$, we observe empty $T=1$ structures at weak $\tilde{\epsilon}_{lj}$ and $R_0/b_0<1.2$ (Fig.~\ref{noneqphase}c). However, for the strong $\tilde{\epsilon}_{lj}$ the most dominant symmetric structures are $ E$ and $E^*$ (Fig.~\ref{noneqphase}d) with hexagonal barrel and tennis ball symmetries, respectively. Both structures have 36 protein subunits. Figure \ref{noneqphase}f shows that while for $R_c/b_0=1.3$ at strong $\tilde{\epsilon}_{lj}$ the structures $J$, $ E$, $ E^*$, $F$ and $F^*$ form, for the weak $\tilde{\epsilon}_{lj}$ several additional symmetric structures form \ref{noneqphase}e. We note that the capsids with the number of subunits less than 36 assemble without packaging the genome because the encapsidation of genomes whose size is much larger than the capsid size is energetically forbidden due to the high cost of elastic energies. 

Figure \ref{noneqphase}g shows the phase diagram for the core size $R_c/b_0=1.4$. For this case, all the capsids with less than 40 subunits are empty. In contrast, Fig.~\ref{noneqphase}h shows that at $R_c/b_0=1.4$, if $\tilde{\epsilon}_{lj}$ is strong, all the capsids encapsidate the genome, and only symmetric structures $ F$ and $ G$ form. As the core size increases to $ R_c/b_0=1.5$, Fig.~\ref{noneqphase}j shows that for strong $\tilde{\epsilon}_{lj}$, the structure $G$ with $T=3$ symmetry covers a large area of the phase diagram. However, at weak $\tilde{\epsilon}_{lj}$, structures with different symmetries form.  

The results presented in Fig.~\ref{noneqphase} clearly show that the genome can completely modify the structure of the capsid. In particular, the structure $F^*$ in Fig.~\ref{noneqphase}d that has not been previously observed in the empty shell simulations, is exactly the structure called H15 obtained in the self-assembly studies of BMV shown in Fig.~\ref{experiments}C and D \cite{bond2020virus}. The structure $E^*$ also covers a broad region of the phase diagram (Fig.~\ref{noneqphase}d), while this structure appears in a tiny region of the phase diagram for the empty shells (Fig.~\ref{empty}). We will discuss these structures and their relation with the experiments \cite{bond2020virus} in the conclusion section.

One of our other intriguing results corresponds to the structure $D$, the so called mini-coat that appears in the assembly studies of clathrin shells too \cite{fotin2004molecular}. Quite interestingly, even when the size of cargo is optimal for the formation of $T=1$ icosahedral structure, depending on the spontaneous curvature of proteins and their mechanical properties, the structure $D$ prevails a large region of phase space (Fig.~\ref{noneqphase}b). This might explain to some extent why minicoats form in the case of clathrin shells, rather than an icosahedral structure. 

As noted above, the white regions of the phase diagrams in Figs.~\ref{empty} and \ref{noneqphase} corresponds to the structures with no specific symmetry.  In the next section, in which we present the equilibrium structures, we will show that the irregular shells in the white regions do never constitute the minimum energy structures. They basically correspond to the kinetically trapped capsids. In fact, all the small shaded regions in the deterministic phase diagrams surrounded by the white region will expand in the equilibrium phase diagrams (see Fig.~\ref{eqphase}) and the white regions will completely disappear. 

Another interesting point of the phase diagrams is that the genome can move the transition point between two icosahedral structures $T=1$ and $T=3$ even when the genome-protein interaction strength is weak ($\tilde{\epsilon}_{lj}=0.005$). For example, at $\gamma= 10$ when there is no core, the transition point from $T=1$ to $T=3$ structure occurs at $R_0/b_0=1.31$ (Fig.~\ref{empty}). However, even when the attraction is weak, for $R_c/b_0=1.5$, the empty $T=1$ forms at very small $R_0/b_0$ without packaging the core since packaging of a big core will cost a lot of elastic energy. Nevertheless, increasing $R_0/b_0$ to 1.24 (Fig.~\ref{noneqphase}i) results into the formation of $T=3$ with the genome enclosed, see Fig.~10 in Appendix.

In general, in the presence of weak LJ, the phase boundaries remain more or less vertical showing that the spontaneous curvature of proteins has the key role in defining the structure of capsids. However, the boundary phases become more horizontal as we increase LJ strength, indicating that the core determines the radius of capsid while FvK number modifies the shell symmetry.

\subsection{Equilibrium structures}
As noted previously, the deterministic simulations involve some irreversible moves, and as such the structures obtained in Fig.~\ref{empty} and Fig.~\ref{noneqphase} do not necessarily constitute the free energy minimum structures. To obtain the equilibrium size and symmetry of empty capsids as a function of $\gamma$ and $R_0/b_0$, in a previous work, we carried out Monte Carlo (MC) simulations combined with the bond-flipping method~\cite{panahandeh2018equilibrium}. Unexpectedly, the equilibrium structure phase diagram and the deterministic one presented in Fig.~\ref{empty} were very similar. It was found that the striking similarities are the result of the strong affinity for the formation of pentamers at specific positions in the shell, which depends on the spontaneous radius $R_0/b_0$ and the mechanical properties ($\gamma$) of the capsid proteins \cite{li2018large,panahandeh2018equilibrium}.
 
Since the previous works have already confirmed that similar structures appear in the deterministic and equilibrium phase diagrams albeit small differences in the boundary between different regions, to obtain the minimum energy structures in the presence of genome, we compared the total energy per subunit of all the structures formed in Fig.~\ref{noneqphase} including the irregular ones for various core sizes as a function of $R_0/b_0$ and $\gamma$. 

To construct the equilibrium phase diagrams, we assumed that the genome is located in the center of mass of the capsid and is interacting with the triangular subunits through the LJ potential. The elastic energy of the capsid as before is the sum of the stretching $E_s$ and bending $E_b$ energies as presented in Eqs.~\ref{Es} and \ref{Eb}. Keeping the genome at the center of mass, we let different capsids relax and then we calculate the total energy including the core-capsid, the stretching and bending energies per subunit to find the minimum energy structures.  
It is worth mentioning that instead of constructing the phase diagrams based on the minimum energy calculations, we could have performed simulations in the grand canonical ensemble framework and monitor the growth of the shells \cite{panahandeh2020virus}. That approach would be useful if we were interested in the impact of the protein concentration, temperature and the strength of hydrophobic interaction between the proteins too. We note that all these parameters can modify the final assembly product, see Ref. \cite{paquay2016energetically}. However, this would have added many more degrees of freedom to the problem and make the interpretation of the results less transparent.  To this end, we assumed that the protein concentration and the strength of the hydrophobic interactions are such that only the competition between the elastic energies, the core size and the strength of core-protein interactions define the final structure of capsids. Thus we constructed the equilibrium phase diagrams based on the strength of LJ interactions and elastic energies. 

\begin{figure}
 \centering
  \includegraphics[width=\linewidth]{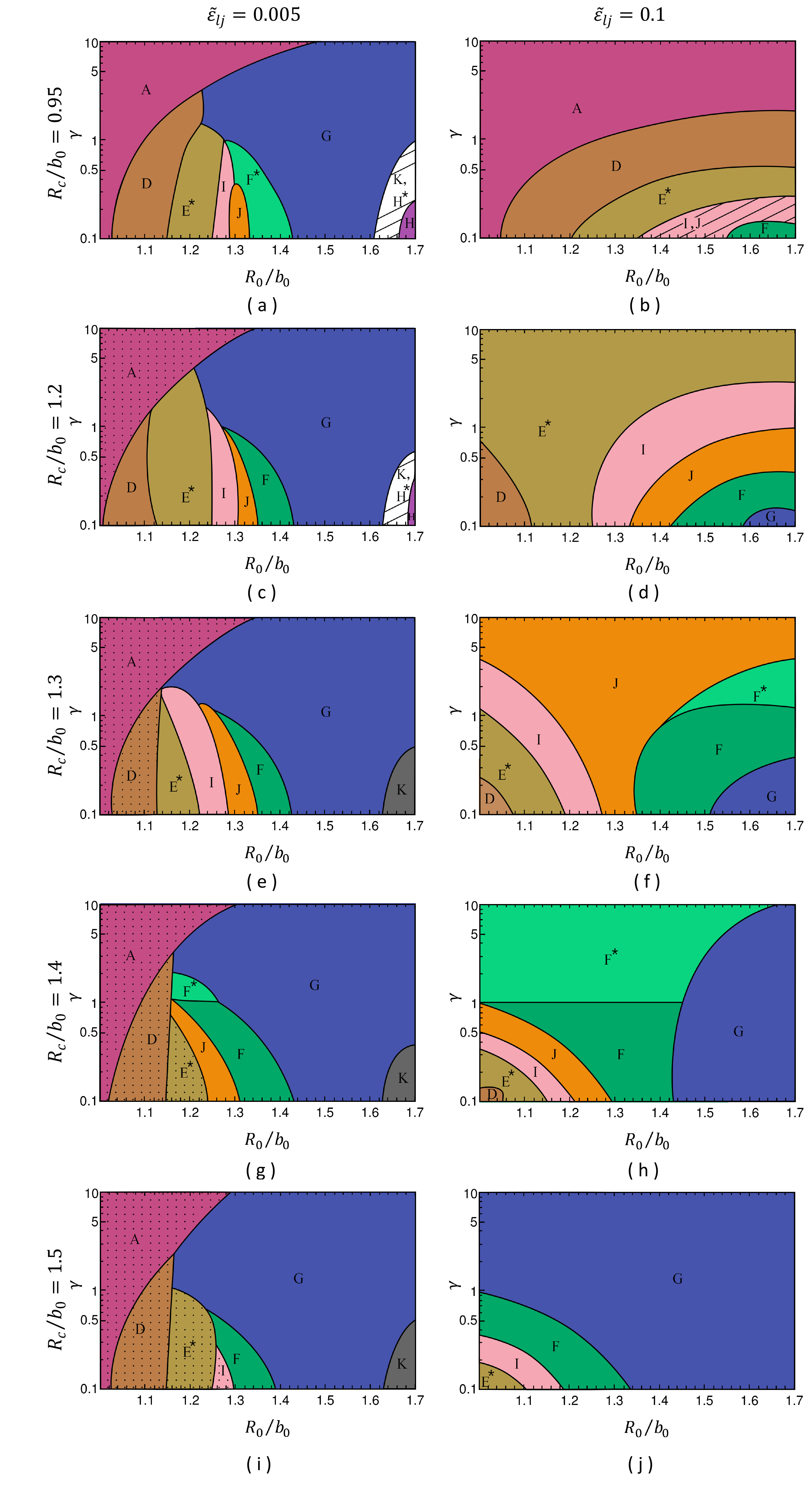}
  \caption{\footnotesize Phase diagrams of the equilibrium structures obtained in the presence of genome. The plots have the same color code as in Fig.~\ref{noneqphase}. The dotted colors correspond to the regions in which the empty shells are the equilibrium structures. From the top to the bottom row, the core sizes are $R_c/b_0=0.95, 1.2, 1.3, 1.4, 1.5$. We note that the core sizes $R_c/b_0 = 0.95, 1.2, 1.4, 1.5$ are the optimal sizes to grow $T=1$, $E$ and $E^*$, $F$ and $F^*$, and $T=3$ structures, respectively. The core size $R_c/b_0 = 1.3$ does not correspond to any of the symmetric structures. The left column corresponds to the weak core-protein interactions ($\tilde \epsilon_{lj}=0.005$) and the right column to the strong one ($\tilde \epsilon_{lj}=0.1$). In the hashed regions, two equilibrium structures form. Each color corresponds to a structure with specific symmetry (see Fig.~\ref{shape}). Structure $J$ is shown in Fig.~11 in the Appendix. Similar to the phase diagrams presented in Fig.~5, the icosahedral structures prevail even if both the spontaneous curvature and core sizes are optimal for the assembly of other symmetric structures.}
  \label{eqphase}
\end{figure}

Figure \ref{eqphase} shows the equilibrium phase diagrams for the same core sizes ($R_c/b_0$) as the ones used in Fig.~\ref{noneqphase}, from top to bottom $R_c/b_0 = 0.95, 1.2, 1.3, 1.4, 1.5$.  
Similar to Fig.~\ref{noneqphase}, the left and right columns correspond to the weak ($\tilde{\epsilon}_{lj}=0.005$) and strong ($\tilde{\epsilon}_{lj}=0.1$) LJ interactions, respectively. In the dotted regions, only empty shells can form. Comparing the equilibrium phase diagrams with the non-equilibrium ones, we find that several structures assembled in the deterministic simulations do not represent the equilibrium structures, such as $B$, $C$ and $E$. A quite interesting difference is that the structure $F^*$, discovered in the experiments of Ref.~\cite{bond2020virus}, appears in wide regions of the phase diagrams in Fig.~\ref{eqphase}. However, the structure $E$ that was also obtained in the recent experiments, only appears in the deterministic simulations performed under non-equilibrium conditions.  Further, all the regions related to irregular shapes in the non-equilibrium phase diagram disappear in the equilibrium ones, most of them are replaced by the structure $J$ with a two fold symmetry (see Fig.~11 in the Appendix). The icosahedral structure $T=3$ occupies a larger area in the equilibrium phase diagrams, in agreement with the observation that many small ssRNA viruses form $T=3$ structures.

\begin{figure}
 \centering
  \includegraphics[width=\linewidth]{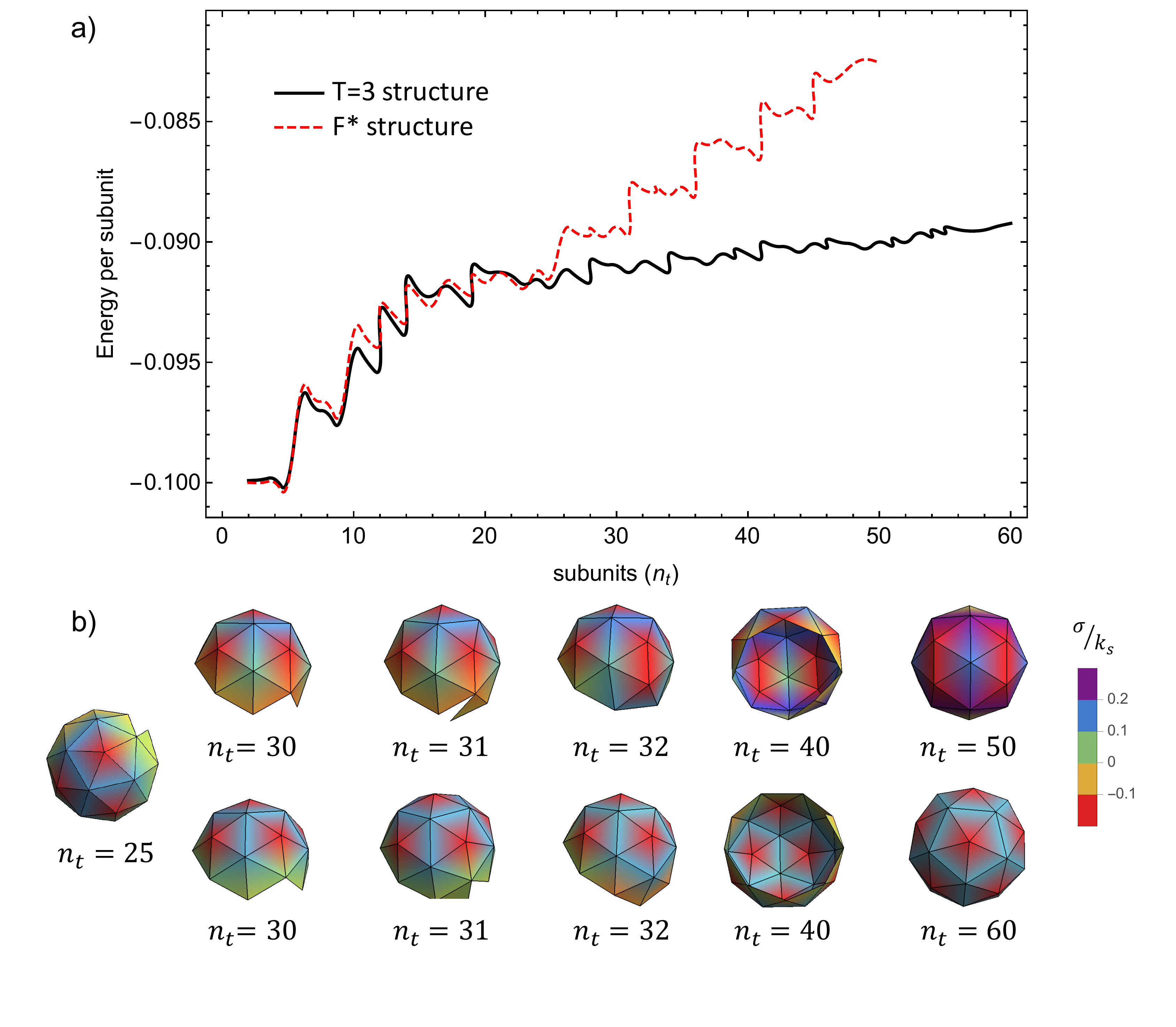}
  \caption{\footnotesize a) Plots of the energy per subunit {\it vs.}~number of subunits for the structures $F^*$ (the red dashed line) and $T=3$ (the black solid line) at $\gamma=5$ and the strong LJ potential $\tilde \epsilon_{ij}=0.1$. The spontaneous radius of curvature of proteins in both structures is $R_0/b_0=1.4$, which is the preferred curvature for the formation of $F^*$. If the size of the core is also $R_c/b_0=1.4$, $F^*$ constitutes the minimum energy structure, as expected.  However, if the size of the core is $R_c/b_0=1.5$, $T=3$ becomes the equilibrium structure despite the fact that the spontaneous radius of proteins is $R_0/b_0=1.4$, which commensurates with an $F^*$ structure. b) Map of the stress distribution in several incomplete shells growing within the pathways shown in a). The top and bottom rows show some intermediate structures of the $F^*$ and $T=3$ shells, respectively. Up to 30 subunits, the intermediate structures of both $F^*$ and $T=3$ look exactly the same.  However, their energies become different after 25 subunits are assembled. The color bar shows the different levels of stress. Violet color corresponds to the maximum of stretching stress and the red one to the maximum of compression. The $F^*$ structure is under much more stress compared to $T=3$. 
  }
  \label{FsT3}
\end{figure}

 To verify the growth pathway suggested in Ref.~\cite{bond2020virus} and described in the introduction, we closely study the assembly of $T=3$ and $F^*$ in the presence of different core sizes.  Figures~\ref{eqphase}h and j show that at high $\gamma$s, for the core size $R_c/b_0=1.4$, $F^*$ is the equilibrium structure; however, increasing $R_c/b_0$ to 1.5, $T=3$ becomes the minimum energy structure in the same region of the phase diagram (for the same $\gamma$ and $R_0/b_0$). Figure~\ref{FsT3}a shows the growth pathways as the plots of the energy per subunit, for $F^*$ and $T=3$ structures, as a function of number of subunits at $R_c/b_0=1.4$ and $R_c/b_0=1.5$, respectively. The other parameters in the figure are the spontaneous radius of curvature $R_0/b_0=1.4$, $\gamma=5$ and $\tilde \epsilon_{ij}=0.1$. The figure clearly shows that up to 25 subunits the energies of the two structures are very close to each other. Note that the total number of subunits in $F^*$ structure is 50 subunits, and thus 25 is when the half of the capsid is formed. From this point to the end of the growth though, the energy of $F^*$ increases sharply in comparison to $T=3$. 
 
 We emphasize that as shown in Figs.~\ref{eqphase}h and j, the optimal spontaneous radius of curvature for the formation of $F^*$ structure is $R_0/b_0=1.4$, explaining why if the core size is also $R_c/b_0=1.4$, $F^*$ is the minimum energy structure. However, if we set the radius of the core to $R_c/b_0=1.5$ while keeping the preferred spontaneous curvature of proteins the same as before ($R_0/b_0=1.4$), then we obtain $T=3$ structures even though $R_0/b_0$ is the optimal size for the formation of $F^*$. The phase diagrams in Figs.~\ref{eqphase}h and j reveal the regions in which an increase in the core size modifies the optimal structure from $F^*$ to $T=3$. Quite interestingly, this indicates that if two different genome sizes are in solution, the capsid proteins energetically prefer to assemble around the larger one to form $T=3$ structure, and this happens, to some extent, regardless of the preferred curvature of proteins.
 
 Figure~\ref{FsT3}b shows the stress distribution of some incomplete structures corresponding to the growth of $F^*$ and $T=3$ shells in Fig.~\ref{FsT3}a. See Appendix for the details of the stress calculation \cite{lizandigrason}. The bar legend on the right side of Fig.~\ref{FsT3}b shows the color code corresponding to the different stress level. While the violet color corresponds to the maximum stretching tension, the red shows the maximum compression stress in the bonds. The first incomplete structure on the left of the figure with $n_t=25$ corresponds to both $T=3$ and $F^*$. Up to that point, the growth pathways of both structures overlap as is illustrated in Fig.~\ref{FsT3}a too. From $n_t=25$ onward, the stress distribution in the structures packaging $R_c/b_0=1.4$ and $R_c/b_0=1.5$ will become different although their growth pathways are still the same {\it i.e.}, their incomplete structures are almost identical regardless of the stress distribution. Finally, at $n_t=30$, the incomplete structures forming around $R_c/b_0=1.4$ and $R_c/b_0=1.5$ start to deviate and follow two different pathways. The upper row shows the pathway corresponding to the formation of the $F^*$ structure with 50 subunits. The figure shows that the highest compression stress distribution is along the pairs of pentamers, where the axis of two fold symmetry are located. However, the bottom row corresponds to $T=3$ with 60 subunits. The level of stress in the $T=3$ structure is much lower than the $F^*$ structure, confirming that $F^*$ can easily break into large fragments to provide the intermediate structures for the formation of $T=3$ with lower stress when cognate larger RNA becomes available.
 
Another puzzling question to address corresponds to as why the structure $F^*$ appears only in a small region in the deterministic simulations but covers large regions in the equilibrium phase diagrams. For example, at $R_0/b_0=1.1$, $\gamma=3$, $R_c/b_0=1.4$ and the strong LJ potential ($\tilde{\epsilon}_{lj}=0.1$), $F^*$ is the minimum energy structure, see Fig.~\ref{eqphase}h. However, in this region, we obtain the structure $F$ in our deterministic simulations as shown in Fig.~\ref{noneqphase}h. A careful review of the step by step assembly of both structures (see Fig.~12) shows that both structures $F$ and $F^*$ are built from 50 subunits, and both follow exactly the same pathway till 25 subunits are assembled (the half of the capsid). 
After the assembly of the first 25 subunits (Fig.~\ref{Fs}a), $F$ and $F^*$ follow two different pathways. For the structure $F$, the five subunits sitting at the edge of the growing shell merge and form a pentamer (Fig.~\ref{Fs}b). However, in the case of $F^*$, one more subunit is added and the capsomer will close as a hexamer (Fig.~\ref{Fs}c). Even though a complete $F^*$ structure has a lower energy than the $F$ one, the assembly of a hexamer is energetically more costly than a pentamer after the half of the capsids is assembled. To this end, there is an energy barrier to the formation of $F^*$ and thus in the deterministic simulations, the shells follow the pathway resulting into the formation of $F$ structures.

\begin{figure}
 \centering
 \includegraphics[width=0.7\linewidth]{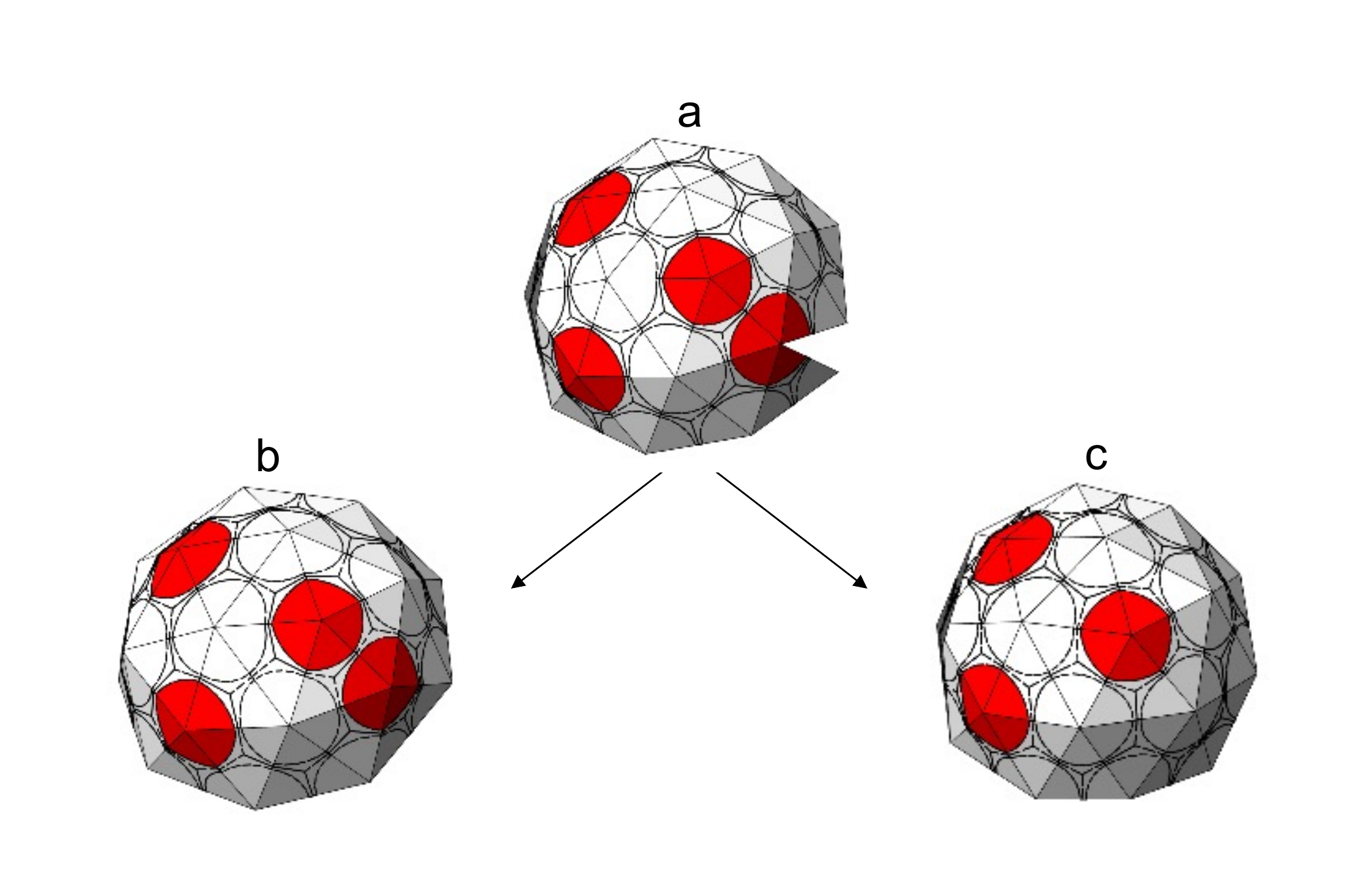}
 \caption{\footnotesize Both structures $F$ and $F^*$ follow the same pathway until the half of the capsids (25 subunits) is assembled (a).  Then, the two structures follow two different pathways.  For the $F$ structure, the 5 subunits at the edge merge and form a pentamer (b). However, for the $F^*$ structure, one more subunit will be added and the capsomer will close as a hexamer (c). The energetic cost of the formation of a hexamer is higher at this stage, and as such, the capsid will follow the $F$ structure pathway while the energy of a complete $F^*$ structure is lower. Red and white colors correspond to the location of pentamers and hexamers respectively. See also Fig.~12 in Appendix for the plot of energy per subunit {\it vs}.~the number of subunits.
  }
 \label{Fs}
\end{figure}

\section{\label{conclusion}Conclusion}

Very little is known as to why inside a cell, capsid proteins predominantly package their cognate RNA. Many factors contribute to this process, but our knowledge in this area is still rudimentary.  In this paper, we have investigated the impact of genome on the structure and excess elastic stress of viral shells, which are tightly connected to the capsid stability and function.  In particular we have focused on the recent self-assembly experiments of BMV capsid proteins around short oligonucleotides, using a combination of charge detection mass spectrometry and cryo-EM techniques \cite{bond2020virus}.  Quite interestingly, these experiments have revealed that the capsid proteins can assemble to form symmetric structures other than icosahedral ones \cite{bond2020virus}. Some of the structures formed around the small pieces of oligonucleotides have the $D_{6h}$ hexagonal barrel symmetry (the $E$ structure in Fig.~\ref{experiments}A and Fig.~\ref{shape}) and some others have $D_{5h}$ symmetry (the $F^*$ structure in Fig.~\ref{experiments}C and Fig.~\ref{Fframes}).  

To explore under what conditions the $F^*$ and $E$ structures could form, we have constructed several phase diagrams for various genome sizes as a function of spontaneous curvature and mechanical properties of protein subunits, see Figs.~\ref{noneqphase} for the results of the deterministic simulations and \ref{eqphase} for the equilibrium structures.  We found that the structure $F^*$, which was called $H15$ in Ref.~\cite{bond2020virus}, can only form in the presence of genome.

Our results also show that the size of genome plays an important role for the assembly of $F^*$, and that the optimal core size for the formation of $F^*$ is smaller than that for $T=3$. As illustrated in the plots of Fig.~\ref{eqphase}h, the shell $F^*$ constitutes the global minimum energy structure at the smaller core size ($R_c/b_0=1.4$) for a large region of the phase diagram. Figure \ref{FsT3} reveals that the formation of $F^*$ and $T=3$ follow almost the same pathway until 30 triangular subunits are assembled.  However, their energies and stress distributions start differing after the assembly of the first 25 subunits. Figure \ref{FsT3}a shows that after the assembly of the first 30 triangular subunits, while $F^*$ is an equilibrium structure for $R_c/b_0=1.4$, if a larger RNA becomes available, the shell can easily switch the pathway and form a $T=3$ structure because of the difference between the energy of a $T=3$ and $F^*$ for larger RNAs. 

We also calculated the stress distribution across $F^*$ and $T=3$ shells as they grow (see Fig.~\ref{FsT3}b), which confirms our conclusion based on the energy of the systems. As can be seen in the figure, the stress between every two adjacent pentamers in $F^*$ structure is very high despite the fact that it is an equilibrium structure, Fig.~\ref{eqphase}h. The implication is that $F^*$s are stabilized by encapsidating short RNAs, but they can become metastable and split into large fragments, which later grow to form $T=3$ structure with much lower stress distribution if larger cognate RNAs are accessible.  Thus, our work suggest an alternative pathway for the assembly of $T=3$ shells inside the cell in the presence of many smaller non-viral RNAs.

Another interesting point of the {\it in vitro} experiments of Bond {\it et al.}~\cite{bond2020virus} corresponds to the co-existence of $E$ and $F^*$ structures. Our work clarifies the conditions under which the two different structures can coexist.  Since the capsids assembled in the experiments are built from the same types of proteins, we explored if it is possible to obtain both structures ($E$ and $F^*$) for the same $\gamma$ and $R_0/b_0$ (related to the mechanical properties of capsid proteins), albeit different core sizes.  A careful review of Figs.~\ref{noneqphase} and \ref{eqphase} shows that we can obtain several structures simultaneously in various cases if different genome lengths are packaged inside the capsids.  These results are consistent with the findings of the experiments of Ref.~\cite{bond2020virus}, which show that the number of packaged short oligonucleotides are not the same in each structure.  Thus if the number of packaged oligomers are different, it is possible to obtain the structures $E$ and $F^*$ simultaneously in solution.  

It is important to note that the $E$ structure can only be obtained in the deterministic simulations; it never appears in the equilibrium phase diagrams, and as such it constitutes a minimum energy structure neither for the filled nor for the empty capsids.  The $E$s are probably trapped in a local minimum free energy and are less stable than other structures in the experiments, see also the stress distribution across the $E$ structure in Fig.~13.  These structures might form in the cell but can easily disassemble and reassemble to form a stable icosahedral structure.  It is worth mentioning that our simulations were performed under the assumption that the genome assumes an isotropic shape inside the capsid. However, the structure $E$ is slightly elongated and an anisotropic genome distribution inside the capsid can render $E$ a minimum energy structure. Thus, more theoretical and simulations are needed to study the stability of $E$ structures. 

In summary, our work shows that the interplay of protein mechanical properties, genome volume, and strength of the genome-capsid protein interaction can significantly modify the symmetry, structure, and stability of the assembly products, and hence kinetic pathways of assembly. Further, our work also supports the presence of an alternative pathway for the formation of $T=3$ icosahedral structures around their cognate RNAs in the presence of oligonucleotides. More experiments to confirm the existence of this pathway would be required, at different salt concentrations, and genome lengths. Such studies are needed in order to understand ssRNA virus assembly in presence of polyanionic, non-cognate potential encapsulation competitors.

\section{\label{Method}Method}

\begin{figure} 
   \centering
    \includegraphics[width=0.8\linewidth]{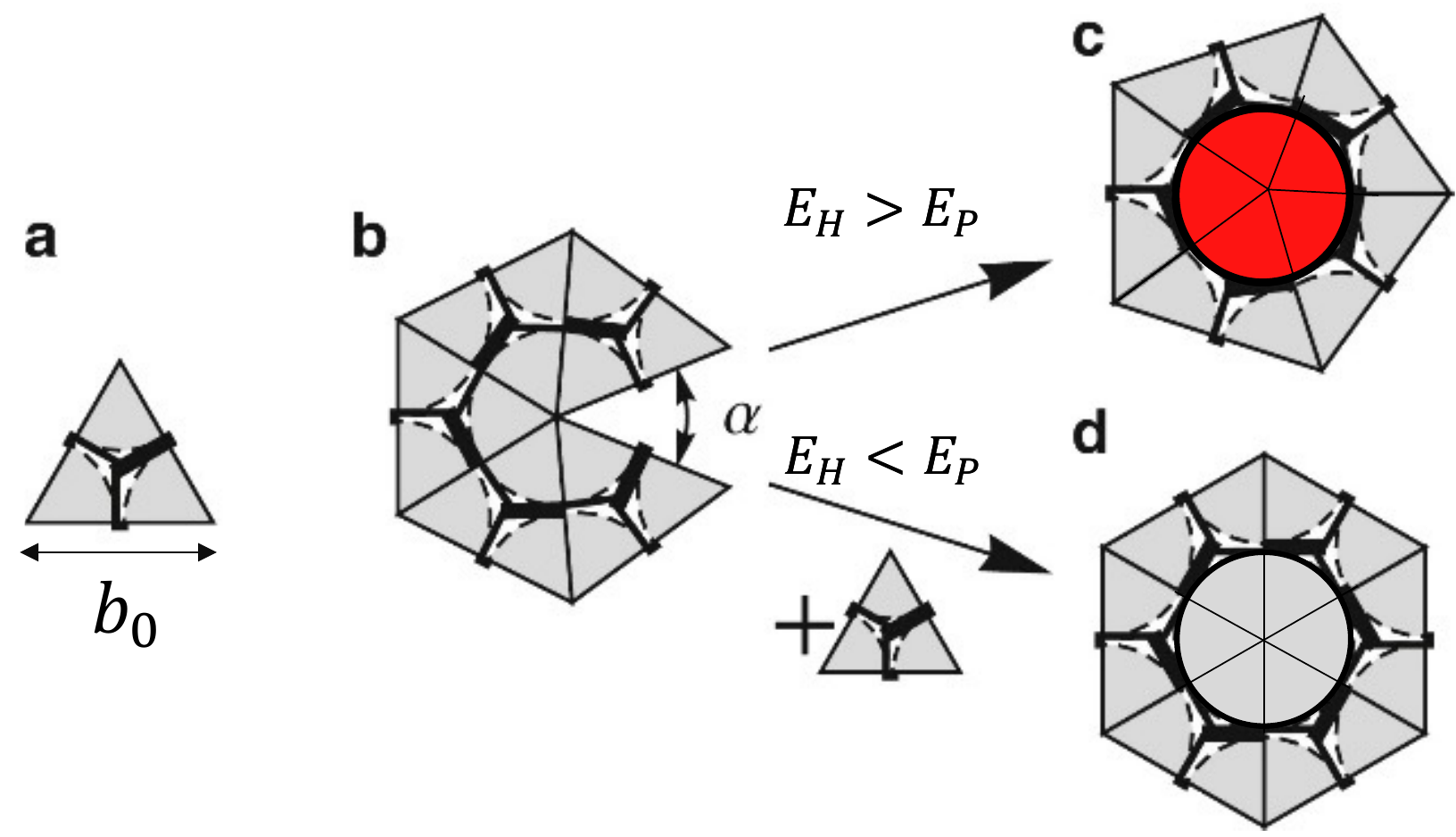} 
    \caption{\footnotesize (a) A protein subunit built from three monomers illustrated as an equilateral triangle in its equilibrium shape.  As triangular subunits are added to the growing shells, because of the curvature of the shell, the subunits can be stretched or compressed. (b) The subunits bind together edge to edge, and the growth of the shell proceeds by adding a subunit to the location with the smallest opening angle, $\alpha$. As the shell grows, the two unbound edges can either bind together to form a pentamer (c), or a new subunit can be added to form a hexamer (d). The choice between forming a pentamer and a hexamer is based on which leads to a lower energy per subunit in the growing shell. If the energy per subunit becomes lower for the formation of a hexamer ($E_H$) compared to that of a pentamer ($E_P$), then a hexamer forms; otherwise a pentamer assembles. Adapted with permission from \cite{wagner2015robust}.}
    \label{fig:subunits}
\end{figure}

To study the formation of capsids around different genome sizes, we use triangular subunits as our building blocks and associate a monomer to each triangle vertex (Fig.~\ref{fig:subunits}a). Since capsids are primarily built from pentamers and hexamers, triangular subunits (trimers) have been widely employed to study the assembly of viral shells and other protein nanocages \cite{vernizzi2011platonic,wagner2015robust,Hagan2013,li2018large}.  The elastic energy of a growing capsid, thus becomes, the sum of the stretching $E_s$ and bending $E_b$ energies as follows \cite{hicks2006irreversible},
the stretching energy results from the deformation of the triangles from their equilateral shape and sums over all bonds $i$. Assuming that each bond is built from a linear spring with the equilibrium length $b_0$ and the spring constant $k_s$, the stretching energy can be written as, 
\begin{eqnarray}\label{Es}
  E_s&=&\sum_{i} \frac{1}{2}k_s(b_{i}-b_0)^2
  \end{eqnarray}
the length of the {\it i}th bond is denoted with $b_i$ and can be stretched or compressed as the shell grows.
  
  The bending energy is due to the deviation of local curvature from the preferred one (to be defined as a parameter in this study) and is calculated by summing over all neighboring pairs of subunits with $<m,n>$ indexing pairs of adjacent subunits and $k_b$ being the bending stiffness as follows,
  \begin{eqnarray}\label{Eb}
  E_b&=&\sum_{<m,n>}k_b[1-\cos(\theta_{<m,n>}-\theta_0)].
  \end{eqnarray}
 The spontaneous radius of curvature of proteins $R_0$ is related to the preferred angle between two subunits $\theta_0$ through the relation $\sin \theta_0/2=(12R_0^2/b_0^2-3)^{(-1/2)}$.
 
 The total elastic energy of the growing shell then depends on two dimensionless parameters, $\gamma=k_s b_0^2/k_b$ and $R_0/b_0$. The Foppl von Karman (FvK) number $\gamma$ indicates the relative difficulty of deforming a subunit from its equilateral triangular shape {\it vs}.~the difficulty of bending it away from the preferred radius of curvature. Both $\gamma$ and $R_0/b_0$ depend on the intrinsic property of proteins (shape and resistance to deformation), and also on the solution circumstances such as pH and ionic strength, and for closed cages also on differentials of osmolyte concentration\cite{fejer2010emergent, zeng2021virus}.
 
To grow a capsid, we consider that the assembly follows the locally minimum energy path, which, under many circumstances, is reasonably close to the most probable growth pathways and thus can yield a meaningful representation of the final assembly products. During the simulations at each growth step a new subunit is added to the growing edge of the incomplete capsid at a position in which the number of neighbors at the vertices of the newly added subunit will be maximized. Hence, the new subunit will be added to a position with the smallest opening angle $\alpha$, see Fig.~\ref{fig:subunits}b.  A defining step in the assembly process corresponds to the formation of pentamers illustrated in Fig.~\ref{fig:subunits}. If there are already 5 triangles attached to a vertex on the rim of an incomplete capsid, then the growth can proceed in two different pathways: (i) a pentamer can be formed by attaching the two neighboring edges, see Fig.~\ref{fig:subunits}c, or (ii) a hexamer can be assembled by inserting a new subunit, Fig.~\ref{fig:subunits}d.  Depending which structure has a lower energy per subunit in the growing shell, a pentamer or a hexamer forms \cite{ning2016vitro}.

Note that the algorithm discussed above for the virus growth corresponds to the physical situation in which the line tension of incomplete capsid is high or subunit-subunit interaction is weak enough such that the proteins can easily move around to explore the energy landscape and then attach themselves to a position where their number of neighbors are maximized \cite{Levandovsky2009}.  Furthermore, we allow the shell to relax between the addition of every subunit assuming that the relaxation time of the elastic stresses in the incomplete capsid is shorter than the time it takes for the shell to grow. This is compatible with the assumption of the high line-tension or weak protein-protein interaction \cite{li2018large, zandi2006classical}. This approach is also consistent with the results of Ref. \cite{elrad2008mechanisms}, in which the self-assembly simulations of $T=1$ structures at different protein concentrations show that slowing the adsorption of free subunits into the core decreases frustration \cite{elrad2008mechanisms}. 

We emphasize that our simulations are deterministic and thus irreversible, mimicking the situation in which the strength of protein-protein interaction is such that once a pentamer or hexamer forms, it can no longer dissociate \cite{ning2016vitro, Levandovsky2009}. Quite interestingly, in our simulations, we do not see aberrant particles as observed in the assembly of $T=1$ structures in Ref. \cite{elrad2010encapsulation}. This might be due to the fact that we follow the most probable pathway \cite{Levandovsky2009}.

To find the locations of high stress points across the capsid, we calculate the elastic stress distribution in the shell through the trace of the stress tensor as follows
\begin{eqnarray}\label{stress}
  \sigma_{v_\alpha}&=&\sum_{v_\beta}(\frac{k_s}{2A}) r_{<v_\alpha,v_\beta>} (r_{<v_\alpha,v_\beta>}-b_0)
  \end{eqnarray}
where the summation is over all vertices sharing a bond with the vertex $v_\alpha$. The $r_{<v_\alpha,v_\beta>}$ is the distance between vertices $\alpha$ and $\beta$ and $A = Z (\frac{1}{3})(\frac{\sqrt3}{4}) b_0^2$ is the average area of each vertex with $Z$ the coordination number.

To model the genome, we consider that viral RNA in solution is believed to be folded in such a way that its radius of gyration only changes by a factor of~3 inside the capsid \cite{yoffe2011ends}. Moreover, several {\it in vitro} virus self-assembly experiments indicate that RNA collapses into its final size right at the beginning of the assembly after a small amount of proteins are adsorbed to it, suggesting that the time scale associated with genome condensation is much faster than the overall time scale for the growth process \cite{Borodavka2012,Chevreuil2018,DanielDragnea2010}. Therefore, we assume that capsid growth occurs in presence of a globular RNA modeled by a spherical core, which reduces considerably the many degrees of freedoms involved in the problem. This will make the assembly process more tractable, allowing us to explore how the shell size and structure depends on the genome size and the subunit-genome interaction. Thus, we consider a spherical core interacting through Lennard-Jones (LJ) with the capsid proteins
  \begin{eqnarray}\label{Elj}
  E_{lj}&=&\sum_{v_\alpha}4 \epsilon_{lj} [(\frac{\sigma}{r_{<v_\alpha,g>}})^{12}-(\frac{\sigma}{r_{<v_\alpha,g>}})^6 ]
  \end{eqnarray}
where $\epsilon_{lj}$ is the potential strength, $\sigma$ is the distance at which the LJ potential between the core and triangle vertex (monomer) is zero and $r_{<v_\alpha,g>}$ is the distance between the core center and triangle vertex $v_\alpha$.

We start the growth process by adding one subunit to the core and then we monitor the growth of the shell.  We assume that the solution condition is always such that the proteins prefer to assemble rather than to stay free in the solution. We consider that the concentration of free proteins in solution is such that the capsid proteins nucleate only in one location along the genome, and the rest of capsid grows around the nucleating site.

In the results and discussion section, we will discuss the conditions under which if the spontaneous radius of proteins is smaller than the size of the core, the shell might not grow around the genome but will form an empty shell.

\begin{acknowledgements}
This work was supported by the National Science Foundation through Grant Numbers DMR-1719550 and 2034794(to R.Z.). B.D. acknowledges support from the Army Research Office, under awards W911NF2010072 and W911NF2010071, and from the National Science Foundation, under award CBET-1803440.
\end{acknowledgements}

\appendix
\section{}
\begin{figure}[h]
 \centering
  \includegraphics[width=1\linewidth]{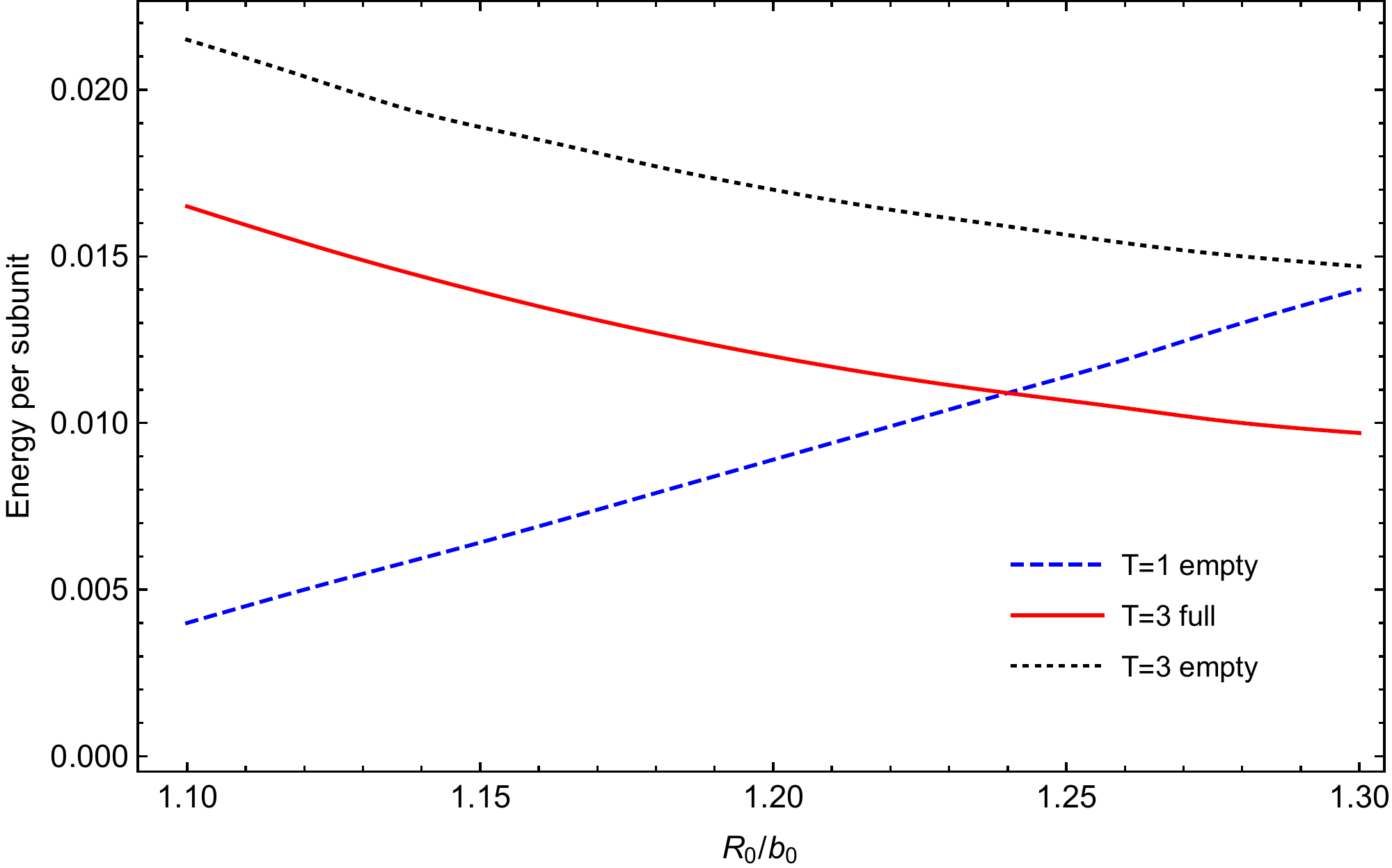}
  \caption{\footnotesize At $\gamma= 10$ when there is no core, the transition point from $T=1$ to $T=3$ is at $R_0/b_0=1.31$. However, in the presence of a core whose size is $R_c/b_0=1.5$ (perfect for a $T=3$ shell), the boundary shifts towards smaller $R_0/b_0=1.24$. When the core-protein attraction is weak ($\tilde{\epsilon}_{lj}=0.005$), $T=1$ forms for small values of $R_0/b_0 < 1.24$ without packaging the core as it will cost a lot of elastic energy. However, increasing $R_0/b_0$ to 1.24, $T=3$ structures form with the genome inside.}
  \label{T1T3R0}
\end{figure}

\begin{figure}
 \centering
  \includegraphics[width=0.3\linewidth]{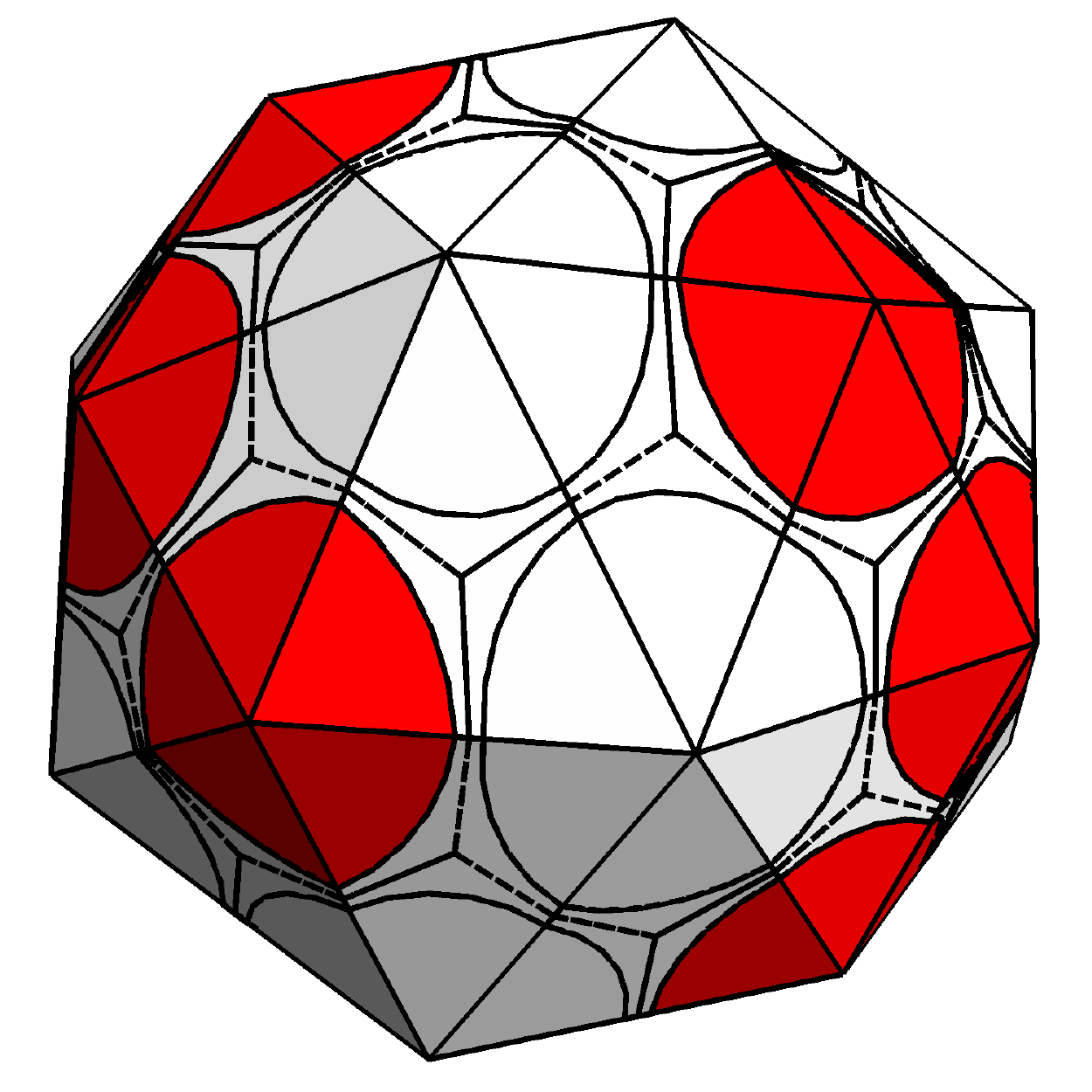}
  \caption{\footnotesize Structure $J$ with 44 subunits and $D_2$ syemmetry. Reproduced from Ref.\cite{panahandeh2018equilibrium} with permission from the Royal Society of Chemistry.}
  \label{Jstruct}
\end{figure}

\begin{figure}
 \centering
 \includegraphics[width=1\linewidth]{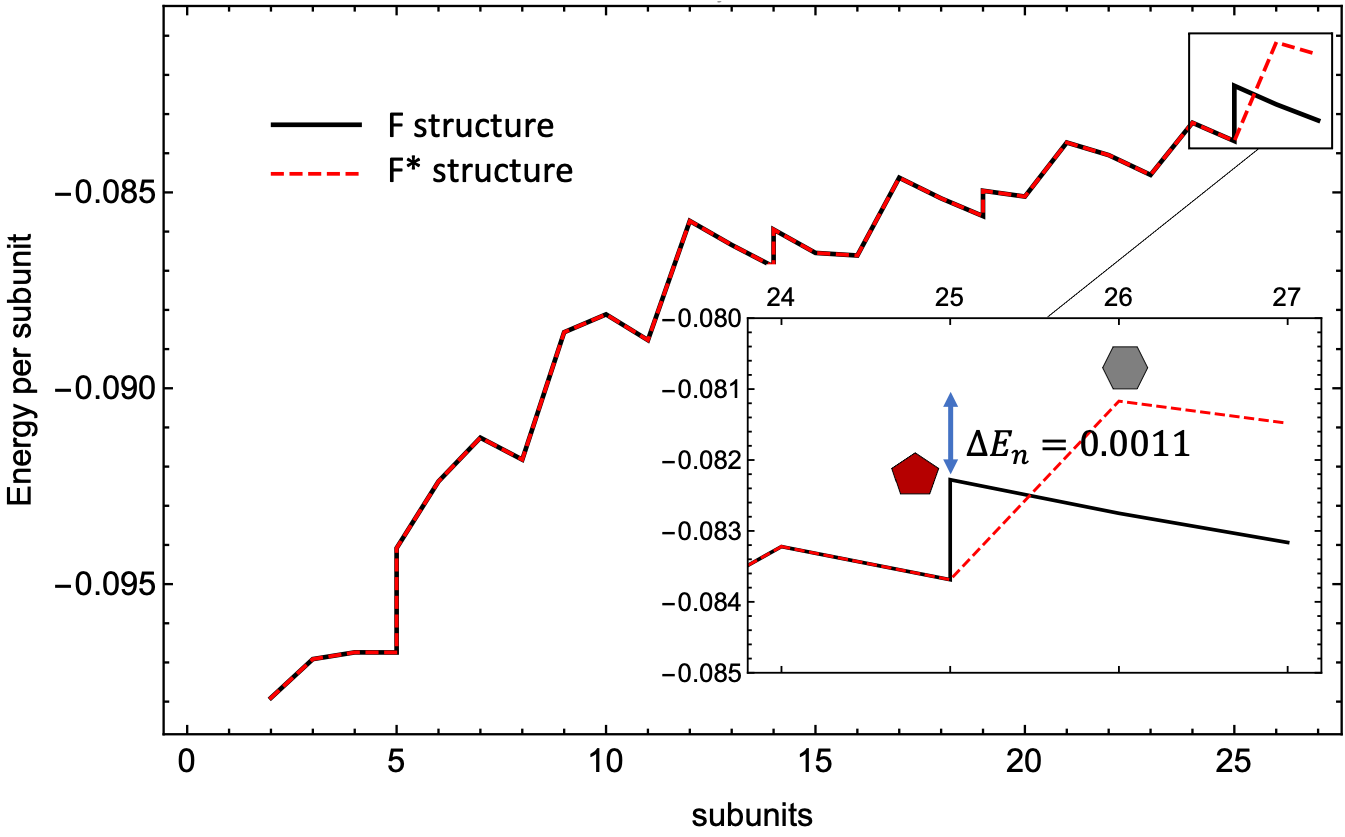}
 \caption{\footnotesize Plots of the energy per subunit for the structures $F$ (the black solid line) obtained from the deterministic simulations and $F^*$ (the red dashed line) as a function of the number of subunits at $R_0/b_0=1.1$, $\gamma=3$, $R_c/b_0=1.4$ and the strong LJ potential $\tilde \epsilon_{ij}=0.1$.  Both structures $F$ and $F^*$ follow the same pathway until one half of the capsids (25 subunits) are assembled. Then, the two structures follow two different pathways.  For the $F$ structure, the 5 subunits at the edge merge and form a pentamer. However, for the $F^*$ structure, one more subunit will be added and the capsomer will close as a hexamer with higher energy (see Figure 8 in the paper). The energetic cost of the formation of a hexamer is higher at this stage, and as such, the capsid follows the black line pathway and forms the $F$ structure while the energy of complete $F^*$ structure is lower.
 }
 \label{Fs}
\end{figure}

\begin{figure}
 \centering
  \includegraphics[width=0.8\linewidth]{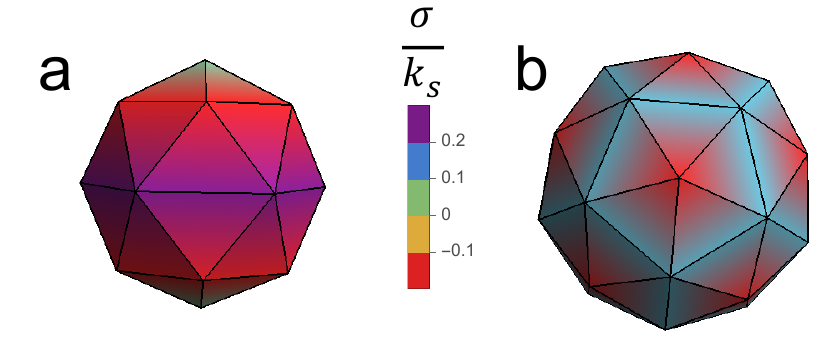}
  \caption{\footnotesize Map of the stress distribution of the structures $E$ and $T=3$ (a) Shell with $E$ symmetry forms in the deterministic simulations for $R_c/b_0=1.2$, $\tilde \epsilon_{lj}=0.1$, $R_0/b_0=1.04$ and $\gamma=3$. The highest positive stress is along the hexamer belt in the middle of the shell. (b)  When we increase the size of the core to $R_c/b_0=1.5$ keeping all the other parameters as in part (a), the structure $T=3$ forms which has much lower stress compared to the structure $E$. 
  }
  \label{stress2}
\end{figure}

\subsection{stress}
In this section, we provide the details of the stress calculation used in Fig.~7 in the paper and Fig.~12 in Appendix. Employing the virial formula, we can calculate the stress component for vertex $v_j$ as follows,
\begin{equation}\label{eq_sigma}
    \sigma^{ij}_{v_\alpha}=\frac{1}{2A}\sum_{v_\beta}(x^i_{v_\beta}-x^i_{v_\alpha}) f^j_{<v_\alpha, v_\beta>},
\end{equation}
where $A=Z \frac{1}{3}\frac{\sqrt{3}}{4}b_0^2$ is the area of vertex $v_\alpha$ with $Z$ the coordination number and $x^i_{v_\alpha}$($x^i_{v_\beta}$) is the $i^{th}$ component of the position of vertex $v_\alpha$($v_\beta$). The summation is over the nearest neighbor vertices. The quantity $f^j_{<v_\alpha, v_\beta>}$ is the $j^{th}$ component of the force on vertex $v_\alpha$ in the $<v_\alpha, v_\beta>$ bond and is equal to
\begin{equation}\label{eq_f}
    f^j_{<v_\alpha, v_\beta>}=k_s(r_{<v_\alpha, v_\beta>}-b_0)\frac{x^j_{v_\beta}-x^j_{v_\alpha}}{r_{<v_\alpha, v_\beta>}},
\end{equation}
where $k_s$ denotes the stretching modulus, $r_{<v_\alpha, v_\beta>}$ is the distance between $v_\alpha$ and $v_\beta$, and $b_0$ is the equilibrium length of the bond. Combining Eqs.~\ref{eq_sigma} and \ref{eq_f}, we obtain
\begin{equation}
    \sigma^{ij}_{v_\alpha}=\frac{k_s}{2A}\sum_{v_\beta} (r_{<v_\alpha, v_\beta>}-b_0)\frac{(x^i_{v_\beta}-x^i_{v_\alpha})(x^j_{v_\beta}-x^j_{v_\alpha})}{r_{<v_\alpha, v_\beta>}}.
\end{equation}
The trace of the stress tensor thus becomes,
\begin{equation}
    \sigma^{kk}_{v_\alpha}=\frac{k_s}{2A}\sum_{v_\beta} r_{<v_\alpha, v_\beta>}(r_{<v_\alpha, v_\beta>}-b_0).
\end{equation}

\nocite{*}

\bibliography{apssamp}

\end{document}